\documentclass[12pt]{article}
\def\lsim{\raisebox{-4pt}{$\,\stackrel{\textstyle{<}}{\sim}\,$}}

\newcommand{\zpr}{\mbox{$Z^{\prime}$}}
\newcommand{\upr}{\mbox{$U(1)^{\prime}$}}

\newcommand{\mz}{\mbox{$M_Z$}}
\newcommand{\mw}{\mbox{$M_W$}}
\newcommand{\x}{\mbox{$\times$}}

\begin{document}
\begin{flushright}
\baselineskip=12pt
UPR-0983T \\
FT 2002-03 \\
MADPH-02-1266 \\
\end{flushright}

\begin{center}
\vglue 1.5cm
{\Large\bf The $Z-Z^{\prime}$ Mass Hierarchy in  a Supersymmetric 
Model with a Secluded $U(1)^{\prime}$-Breaking Sector}
\vglue 2.0cm
{\Large Jens Erler$^{1,2}$, Paul Langacker$^{1,3}$ and Tianjun Li$^{1}$}
\vglue 1cm
{$^1$ Department of Physics and Astronomy \\
University of Pennsylvania, Philadelphia, PA 19104-6396 \\  
$^2$ Universidad Nacional Aut\'onoma de M\'exico \\
Instituto de F{\'\i}sica,  01000 M\'exico D.F. \\
$^3$ Department of Physics, University of Wisconsin, Madison WI 53706}
\end{center}

\vglue 1.0cm
\begin{abstract}

We consider the $\zpr/Z$ mass hierarchy in a supersymmetric model in
which the \upr \ is  broken in a secluded sector coupled to the ordinary
sector only by gauge and possibly soft terms. A large mass hierarchy
can be achieved while maintaining the normal sparticle spectra if there
is a direction in which the tree level potential becomes flat when a
particular Yukawa coupling vanishes. 
We describe the conditions needed for the desired breaking
pattern, to avoid unwanted global symmetries, and
 for an acceptable effective $\mu$ parameter.
The electroweak breaking is dominated by $A$ terms rather than scalar
masses, leading to 
$\tan\beta \simeq 1$.
The spectrum of the symmetry breaking sector is displayed. There is significant
mixing between the MSSM particles and new standard model singlets, for both the
Higgs scalars and the neutralinos.
A larger Yukawa coupling for the effective $\mu$ parameter is 
allowed than in the NMSSM because of the \upr \ contribution to
the running from a high scale.
The upper bound on the tree-level mass of
 the lightest CP even  Higgs doublet mass
is about $c \times $174 GeV, where $c$ is of order unity,
but the actual mass eigenvalues
 are generally smaller because of  singlet mixing.

\end{abstract}

\vspace{0.5cm}
\begin{flushleft}
\baselineskip=12pt
\today\\
\end{flushleft}
\newpage
\baselineskip=14pt

\section{Introduction}

The possibility of an extra \upr \ gauge symmetry 
is well-motivated in superstring constructions~\cite{string}
and grand unified theories~\cite{review},
and also in  models of dynamical symmetry breaking~\cite{DSB}.
In supersymmetric models, an extra \upr \ can provide an elegant
solution to the $\mu$
problem~\cite{muprob1,muprob2}, with an effective $\mu$ parameter
generated by the vacuum expectation value (VEV) of the Standard Model (SM) singlet
field $S$ which breaks the \upr \ symmetry. This is somewhat similar to
the effective $\mu$ parameter in the Next to Minimal Supersymmetric Standard
Model (NMSSM)~\cite{NMSSM}. However, with a \upr \ the extra discrete
symmetries and their associated cosmological problems typically associated
with the NMSSM are absent. A closely related feature is that the MSSM
upper bound of \mz \ on the tree-level mass of the
corresponding lightest MSSM Higgs scalar
is relaxed, both in models with a \upr \ and in the NMSSM, because of 
the Yukawa term $h S H_1 H_2$ in the superpotential and the
$U(1)'$ $D$-term~\cite{Higgsbound}.
More generally, for specific \upr \ charge assignments for
the ordinary and exotic fields one can simultaneously ensure the absence of
anomalies; that all fields of the TeV-scale
effective theory are chiral, avoiding a generalized
$\mu$ problem; and the absence of dimension-4 proton decay
operators~\cite{general}.

In superstring-motivated models it is often
the case that electroweak and \upr \ breaking are both driven by
soft supersymmetry breaking parameters, so one typically expects the \zpr \ mass
or masses
to be of the same order as the electroweak scale\footnote{One way to avoid
this conclusion is for the \upr \ breaking to occur along a direction
which is $F$ and $D$ flat at the renormalizable tree level~\cite{intscale}.},
i.e., less than a TeV or
so~\cite{string}, so that such particles, if they exist,  should be easily 
observed and their couplings studied at future colliders or at the
Tevatron~\cite{collider}. 
The typical expectation is that the \zpr \ mass should be comparable
to \mw \ and \mz. However, there are stringent limits from direct searches
during Run I at the Tevatron~\cite{explim} and from indirect precision tests at the
$Z$-pole, at LEP 2, and from weak neutral current experiments~\cite{indirect}.
The constraints depend on the particular \zpr \ couplings, but in typical
models one requires $M_{Z^{\prime}} > (500-800) $ GeV
and the $Z-\zpr$ mixing angle
$\alpha_{Z-Z^{\prime}}$ to be smaller than a few $\x 10^{-3}$.
(There are actually hints  of deviations from
the standard model in the NuTeV
experiment~\cite{NuTeV} and in atomic parity violation~\cite{APV},
which could possibly be early signs of a \zpr~\cite{hints}.)
The non-observation to date of a \zpr \ 
reduces the attractiveness of such scenarios, but does not exclude them.
It has been shown in a number of examples~\cite{examples} that there are small but
not overly tuned corners of parameter space which can yield acceptable \zpr \
parameters. The most common situation is that the soft-supersymmetry breaking parameters with
dimensions of mass, and therefore the VEV of the field $S$ which
breaks the \upr, are large compared to the electroweak scale, e.g., of $O({\rm TeV})$.
The values of the Higgs doublet 
VEVs, and therefore $M_{W,Z}$, are relatively
small by accidental cancellations. Since the
SUSY-breaking scale is large in such scenarios, they typically lead to nonstandard
sparticle spectra, with heavier squarks and sleptons than for most of the MSSM parameter
space, but a richer spectra of Higgses, neutralinos, and
usually charginos~\cite{altspectrum}.
Another possibility~\cite{examples} is that the electroweak and
\upr \ breaking are driven
by $A$ terms  that are relatively
large compared to the typical soft scalar mass scale. 
This can lead to a small $\alpha_{Z-Z^{\prime}}$ and also a small \zpr ~mass. The latter
might be acceptable if the \zpr \ has strongly suppressed
 couplings to leptons (perhaps after taking
kinetic mixing~\cite{kinetic} into account).

In this paper we consider another possibility, in which all of the
dimensional SUSY-breaking parameters are at or below
the electroweak scale, as is
the VEV of the field which generates the effective $\mu$ term.
Thus, the squark and slepton spectra can mimic those of the MSSM.
The electroweak breaking is actually driven by electroweak scale $A$
terms, with the Higgs doublet and singlet masses smaller.
A large \zpr \ mass can be generated by the VEVs of additional
SM singlet fields that are charged under the \upr.
If these fields are only weakly coupled to the SM fields, i.e.,
by \upr \ interactions and possibly soft SUSY-breaking terms, then
the scale of VEVs in this sector is only weakly linked to the
electroweak scale. In particular, we consider the situation in which
there is an almost $F$ and $D$ flat direction involving these
secluded fields, with the flatness lifted by a small Yukawa coupling. For
a sufficiently small value for this Yukawa coupling, the \zpr \ mass can
be arbitrarily large. The class of models considered is 
related to the intermediate scale models considered in~\cite{intscale},
except that in the latter case the flatness was lifted by
higher dimensional operators or  radiative corrections.

We choose the \upr \ charges so that off-diagonal
soft  supersymmetry breaking mass-square terms can avoid 
unwanted global symmetries, and
show that there are only three  such models up to charge conjugation.
We describe two of these in detail, paying special
attention to avoiding unphysical minima and runaway directions.
Within our assumption of no special adjustment of parameters to
achieve a moderate hierarchy in the ordinary sector, we find
that the Yukawa coupling  associated with the effective $\mu$ parameter
must be relatively large, i.e., of
$O(0.5-0.8)$. The upper end of this range  would lead to a Landau pole
in the NMSSM if one required the theory to be valid up to a
large unification scale~\cite{landau}, but is acceptable in
the \upr \ model due to the new contributions to the renormalization
group equations. (It would be acceptable in either case if one
did not require a canonical desert, as in
models with large extra dimensions.)
This scenario typically generates $\tan \beta \sim 1$, where
$\tan \beta$ is the usual ratio of Higgs doublet VEVs;
that the VEV of $S$ is comparable to that of the doublets; and
that  the upper bound on
 the lightest CP even Higgs doublet 
tree-level mass is of order 170 GeV. However, the actual mass eigenvalues
are reduced by mixing with $SU(2)$ singlets.
For these models, we display the spectra associated with the
symmetry breaking, { i.e}., the gauge bosons, Higgses, neutralinos,
and charginos. There is significant mixing between the standard model
particles and SM singlets in the Higgs and neutralino sectors.
We do not attempt to embed
 the models in a full theory or speculate on the small Yukawa
couplings  needed
either for the large \zpr \ mass or for fermions other than the $t$
quark. In the next section, we discuss the general features of this class of
models.
In Section \ref{models} we calculate the spectra for typical \upr \ charges
and parameter values. Our conclusions are given in
Section \ref{discussion}. Details of the minimization and Higgs
mass-square matrices for one model are given in Appendix A,
and the eigenvectors for the symmetry breaking sector are displayed
for one model in Appendix B.

\section{The Chiral 
Supersymmetric $SU(3)_C \times SU(2) \times U(1) \times U(1)^{\prime}$ Model}
We consider the supersymmetric 
$SU(3)_C \times SU(2) \times U(1) \times U(1)^{\prime}$ model
with 2 Higgs doublets ($H_1$ and $H_2$) and
4 Higgs singlets ($S$, $S_1$, $S_2$, and $S_3$). 
The superpotential  
is\footnote{One might consider a model with 3 
singlets; for example, one can identify 
$S$ with $S_1$. The problem  is that the $F$-term of $S$ is then
$h H_1 H_2 + \lambda S_2 S_3$.
Depending on the soft parameters, there will either be a runaway
direction for the scalar potential that is unbounded from below,
 an unphysical minimum with one of the Higgs doublet VEVs
vanishing, or a minimum in which the
VEVs of $H_1^0$,
$H_2^0$, $S$, $S_2$ and $S_3$ 
are typically of the same order, preventing
a $Z-Z^{\prime}$ mass hierarchy.}
\begin{eqnarray}
W &=& h S H_1 H_2 + \lambda S_1 S_2 S_3 ~,~\,
\end{eqnarray} 
where the Yukawa couplings $h$ and $\lambda$ are respectively associated
with the effective $\mu$ term and with the runaway direction.
The corresponding $F$-term scalar potential is 
\begin{eqnarray}
V_F &=& h^2 \left( |H_1|^2 |H_2|^2 + |S|^2 |H_1|^2 + |S|^2|H_2|^2\right)
\nonumber\\&&
+\lambda^2 \left(|S_1|^2 |S_2|^2 + |S_2|^2 |S_3|^2 + |S_3|^2 |S_1|^2\right)
~,~\,
\end{eqnarray} 
The $D$-term scalar potential is 
\begin{eqnarray}
V_D &=& {{G^2}\over 8} \left(|H_2|^2 - |H_1|^2\right)^2 
\nonumber\\&&
+{1\over 2} g_{Z'}^2\left(Q_S |S|^2 + Q_{H_1} 
|H_1|^2 + Q_{H_2} |H_2|^2 + \sum_{i=1}^3 Q_{S_i}
|S_i|^2\right)^2 ~,~\,
\end{eqnarray}  
where $G^2=g_1^{2} +g_2^2$; $g_1, g_2$,  and $g_{Z'}$ are the coupling constants for
$U(1), SU(2)$ and $U(1)^{\prime}$; and $Q_{\phi}$ is the $U(1)^{\prime}$ charge
of the field $\phi$.

In addition, we introduce the supersymmetry breaking soft terms
\begin{eqnarray}
V_{soft}^{(a)} &=& m_{H_1}^2 |H_1|^2 + m_{H_2}^2 |H_2|^2 + m_S^2 |S|^2 +
\sum_{i=1}^3 m_{S_i}^2 |S_i|^2
\nonumber\\&&
-(A_h h S H_1 H_2 + A_{\lambda} \lambda S_1 S_2 S_3 + {\rm H. C.})
\,  \label{vsoft}
\end{eqnarray} 

There are six neutral complex scalar fields
and  (in the general case)
four phase symmetries of the scalar potential. Two of these are the
$U(1)$ and \upr \ gauge symmetries, implying two unwanted global symmetries.
These will generally be spontaneously broken, implying two massless Goldstone
bosons. One of these has
large $H_1^0$ and $H_2^0$ components and is clearly excluded by experiment.
The second consists mainly of the $S_i$ \ fields, which couple to
ordinary matter only by \upr.
These are most likely also excluded, though a detailed
investigation is beyond the scope of this paper.
We therefore consider special choices for the \upr \ charges which
allow additional (off-diagonal) scalar mass-square terms which
explicitly break the global symmetries.

 For the models considered, one can  take $A_h$ and
$A_\lambda$ to be positive and the extra mass terms added to break the global symmetries
to be negative by an appropriate redefinition of the scalar fields, without loss of
generality.  Then all of the  VEVs can be taken  to be real
and positive at the minima.  We define
\begin{eqnarray}
\langle H_1^0 \rangle \equiv v_1~,~ \langle H_2^0\rangle
 \equiv v_2~,~ \tan\beta={{v_2}\over v_1}
~,~\,
\end{eqnarray}   
and
\begin{eqnarray}
\langle S\rangle \equiv s~,~ \langle S_i\rangle \equiv s_i
~.~\,
\end{eqnarray} 
Note that  we have defined these VEVs without pulling out a
factor of $1/\sqrt{2}$, so the observed value of the electroweak
scale is $\sqrt {v_1^2 + v_2^2} \simeq 174$ GeV. 
We also introduce
\begin{eqnarray}
\Delta \equiv Q_S s^2 + Q_{H_1} v_1^2 + Q_{H_2} v_2^2 + \sum_{i=1}^3 Q_{S_i} s_i^2
~.~\,
\end{eqnarray}

The expressions for the chargino, neutralino, and
 $Z-Z'$ mass matrices are independent of the forms of the
supersymmetry breaking soft terms.
The $Z-Z'$ mass matrix is
\begin{eqnarray}
M_{Z-Z'} =\left(\matrix{M_{Z}^2 & M_{Z Z'}^2\cr
M_{Z Z'}^2 &  M_{Z'}^2\cr}\right) ~,~ \,
\end{eqnarray}
where
\begin{eqnarray}
M_{Z}^2 = {G^2\over 2} (v_1^2 +v_2^2)
~,~ M_{ Z'}^2 = 2 g_{Z'}^2 
\left(  Q_S^2 s^2 + Q_{H_1}^2 v_1^2 + Q_{H_2}^2 v_2^2 + \sum_{i=1}^3 Q_{S_i}^2 s_i^2
\right)~,~\,
\end{eqnarray} 
\begin{eqnarray}
M_{Z Z'}^2 = g_{Z'} G ( Q_{H_1} v_1^2 - Q_{H_2} v_2^2)
~.~\,
\end{eqnarray} 
The mass eigenvalues  are
\begin{eqnarray}
M_{Z_1, Z_2}^2 = {1\over 2} \left(M_Z^2 + M_{Z'}^2 \mp 
\sqrt {(M_Z^2-M_{Z'}^2)^2 + 4 M_{Z Z'}^4 } \right) 
~,~\,
\end{eqnarray} 
and the $Z-Z'$ mixing angle $\alpha_{Z-Z'}$ is given by
\begin{eqnarray}
\alpha_{Z-Z'} = {1\over 2} {\rm arctan} \left({{2 M_{ZZ'}^2}
\over\displaystyle {M_{Z'}^2 - M_Z^2}}\right)
~,~\,
\end{eqnarray} 
which is constrained to be less than a few times $10^{-3}$.

In the basis $\{ {\tilde B}^{\prime}, \tilde B, \tilde W_3^0,
\tilde H_1^0, \tilde H_2^0,
\tilde S, \tilde S_1, \tilde S_2, \tilde S_3\}$, the neutralino mass
matrix is 
\begin{eqnarray}
M_{\tilde \chi^{0}} =\left(\matrix{ M_{\tilde \chi^{0}}(6, 6)
& M_{\tilde \chi^{0}}(6, 3) \cr
M_{\tilde \chi^{0}}(6, 3)^T& M_{\tilde \chi^{0}}(3, 3) \cr}\right)
~,~ \,
\end{eqnarray}
where
\begin{eqnarray}
M_{\tilde \chi^{0}} (6, 6)= \left(\matrix{M_1^{\prime}
&0&0&\Gamma_{H_1}&\Gamma_{H_2}&\Gamma_{S}\cr
0&M_1&0& -{1\over \sqrt 2} g_1 v_1 & 
{1\over \sqrt 2} g_1 v_2 &0\cr
0&0&M_2& {1\over \sqrt 2} g_2 v_1 
& -{1\over \sqrt 2} g_2 v_2&0\cr
\Gamma_{H_1} & -{1\over \sqrt 2} g_1 v_1
 & {1\over \sqrt 2} g_2 v_1
&0&-h s&-hv_2 \cr
\Gamma_{H_2} & {1\over \sqrt 2} g_1 v_2 
& -{1\over \sqrt 2} g_2 v_2
&-h s&0&-h v_1 \cr
\Gamma_S &0&0&-h v_2 & -h v_1 &0 \cr}\right) ,\,
\end{eqnarray}
and
\begin{eqnarray}
M_{\tilde \chi^{0}} (3, 3)= \left(\matrix{ 0& -\lambda s_3&
-\lambda s_2 \cr -\lambda s_3 &0& -\lambda s_1 \cr
-\lambda s_2 & -\lambda s_1 & 0  \cr}\right) ,\,
\end{eqnarray}
where $\Gamma_{\phi} \equiv \sqrt 2 g_{Z'} Q_{\phi} \langle \phi\rangle$;
and $M_1^{\prime}, M_1, M_2$ are gaugino masses for $U(1)^{\prime}$,
$U(1)$ and $SU(2)_L$, respectively.
The first row of $M_{\tilde \chi^{0}} (6, 3)$ is given by
$\left( \Gamma_{S_1}  \ \  \Gamma_{S_2}  \ \ \Gamma_{S_3} \right)$,
while the other entries are zero.

The chargino mass matrix  is
\begin{eqnarray}
M_{\tilde \chi^{\pm}} =\left(\matrix{M_2 & \sqrt 2 M_W ~\sin\beta\cr
\sqrt 2 M_W ~\cos\beta& \mu\cr}\right) ~,~ \,
\end{eqnarray} 
where  $\mu \equiv h s $ is the effective $\mu$ parameter.

If $\mu $ is too small the lighter chargino mass will violate observational
bounds. However, the Yukawa coupling $h$ (at the electroweak scale)
cannot be too
large if the theory is to remain perturbative up to a large grand unification or
string scale. This constraint is somewhat less restrictive than the 
corresponding one in the NMSSM~\cite{landau} because the new contributions from
the \upr \ to the running of $h$ are negative. We have found that
$h$ can be as large as 0.7-0.8, even for $\tan \beta \sim 1$. We will illustrate
the results for the cases $h=0.5$ and $h=0.75$.

For $\lambda\rightarrow 0$ the potential 
may be unbounded below (depending on the $m_{S_i}^2$) for
large $s_i$. In that case, for small but finite $\lambda$ the $s_i$ will
be large, as will the \zpr \ mass. We will typically choose $\lambda$
 to be around $h/10$. Though small, $\lambda$ is still much larger than
most of the Yukawa couplings associated with the fermion masses.

For $A_h$ comparable to the scale of the soft Higgs masses, the potential has
an unwanted global minimum at $v_1 = v_2 = 0$ and $s\sim s_i$. This can be
avoided by choosing $A_h$ to be relatively large, e.g., of order (5-10)
larger than the soft masses. In this case, the symmetry
breaking is driven more by the $A$ terms than the soft masses,
analogous to the large
$A$ scenarios described in ~\cite{examples}. In the large $A_h$ limit
one has $s \sim v_1 \sim v_2$.  For intermediate $A_h$ 
the ratio of $s/v_i$ can be increased to
around $3/2$, but not  much more without introducing the
unwanted minimum described above. 

Therefore,
 the lower bound on the light chargino mass from LEP gives a
strong constraint on the models. This is in the range $\sim(90-104.5)$
for  center mass energy
$\sqrt s=209$ GeV, depending on the decay kinematics~\cite{chargino}.
Let us discuss the chargino masses in detail. They are~\cite{BARTL} 
\begin{eqnarray}
 m_{\tilde \chi^{\pm}_1} &=&
 {\eta_{C1}\over 2} \left(\sqrt {(M_2-\mu)^2 +
 2 M_W^2 (1+\sin2\beta )}
 \right.\nonumber\\&&\left.
 - \sqrt {(M_2+\mu)^2 +
 2 M_W^2 (1-\sin2\beta )}\right) ~,~\, 
\end{eqnarray}  
\begin{eqnarray}
 m_{\tilde \chi^{\pm}_2} &=&
 {1\over 2} \left(\sqrt {(M_2-\mu)^2 +
 2 M_W^2 (1+\sin2\beta )}
 \right.\nonumber\\&&\left.
 + \sqrt {(M_2+\mu)^2 +
 2 M_W^2 (1-\sin2\beta )}\right) ~.~\, 
\end{eqnarray}   
where $\eta_{C1} = \pm 1$ is chosen so  that $m_{\tilde \chi^{\pm}_1}$
is positive. Because $\tan\beta \simeq 1.0$, 
the light chargino mass is
\begin{eqnarray}
 m_{\tilde \chi^{\pm}_1} &\simeq&
 {\eta_{C1}\over 2} \left(\sqrt {(M_2-\mu)^2 +
 4 M_W^2}
 - |M_2 + \mu| \right) ~.~\, 
\end{eqnarray}  
If $M_2$ and $\mu$ have the same sign, {\it i.e.}, $M_2 \mu >0$, this is
typically smaller than ${\rm min} \{M_2, \mu \}$.
For example, for $|\mu| < |M_2|$ and $|M_2| \gg M_W$, one obtains
$m_{\tilde \chi^{\pm}_1} \sim \mu$.
For
 $M_2 \mu < 0$ one finds $m_{\tilde \chi^{\pm}_1} <
\sqrt{\mu^2 + M_W^2}$. The limit is saturated if and only if $M_2=-\mu$, 
in which case the two chargino masses
are equal (for $\tan \beta = 1$). 
Thus, for $h < 0.8$ and $s/v_1\simeq s/v_2 \lsim 3/2$,
 the
upper bounds on the light chargino mass are around  120 GeV and 170 GeV
for the cases $M_2 \mu >0$ and $<0$, respectively, with lower values for
smaller $h$.

The charged Higgs mass is
\begin{eqnarray}
M_{H^{\pm}}^2 = M_W^2 + {{2 A_h h s}\over\displaystyle 
{\sin 2\beta}} - h^2 (v_1^2 +v_2^2) ~,~\,
\end{eqnarray} 
where $M_W^2={g^2\over { 2}} (v_1^2 +v_2^2)$.

The upper bound on the tree-level mass of the  lightest CP even  Higgs
doublet scalar,  i.e.,  before including mixing with the $SU(2)$ singlets
and corresponding to the lightest scalar
 $h^0$ in the MSSM, 
is~\cite{muprob2}
\begin{eqnarray}
m^2_{h^0} \leq M_Z^2 \cos^22\beta + h^2 (v_1^2+v_2^2) \sin^22\beta
+2 g_{Z'}^2 \frac{(Q_{H_1} v_1^2 + Q_{H_2} v_2^2)^2}{(v_1^2+v_2^2)}~.~\, 
\label{doubletlimit}
\end{eqnarray} 
In the models considered here, $\tan\beta \simeq 1.0$, 
so that
\begin{eqnarray}
m_{h^0} \lsim \sqrt {h^2+ \frac{1}{2}g_{Z'}^2(Q_{H_1} + Q_{H_2})^2}
\times 174 \ {\rm GeV}~,~\, \label{mhlim}
\end{eqnarray}
which is much weaker than the corresponding limit
$m_{h^0} < M_Z$ in the MSSM. Of course,
the actual Higgs mass eigenstates involve mixing of the doublets with $SU(2)$
singlets, so that the tree-level mass eigenvalues are lower. Also,
one must add potentially large loop corrections in both cases.
The coefficient $\sqrt {h^2+ \frac{1}{2}g_{Z'}^2(Q_{H_1} + Q_{H_2})^2}$ 
 is typically of order unity.

It is still necessary to have two off-diagonal
dimension-2 soft supersymmetry
breaking terms involving $S$, $S_1$, $S_2$ and $S_3$ to break the two
unwanted global $U(1)$ symmetries. We cannot choose $|Q_{S_1}| \not=
|Q_{S_2}| \not= |Q_{S_3}|$ because then at most one term,
$S S_i$ or $S S_i^{\dagger}$, would be allowed. The only
possibilities are $Q_{S_1} = Q_{S_2} = -{1\over 2} Q_{S_3}$
and $Q_{S_1} =-Q_{S_2}$, $Q_{S_3}= 0$. Any VEV of $S_3$ in the second
case would not be linked to $U(1)^\prime$ breaking; we will not consider
this possibility further.  In the first case, there are two
possibilities for the $U(1)'$ charge of $S$:
$Q_S=\pm Q_{S_1}$ and $Q_S=\pm Q_{S_3}$,
which will be discussed as Model I and Model II,
respectively\footnote{These charges allow additional superpotential terms
$S_1^2S_3$ and $S_2^2S_3$. Their presence would have
little effect on our conclusions, other than changing the relative
sizes of the $s_i$, so they will be ignored for simplicity.
They could also be explicitly  eliminated by discrete
symmetries for the dimension-4 operators, or by string selection rules
if there is an underlying string theory. Similarly, the \upr \ symmetry
would allow additional terms $S S_{1,2}$ (Model I) or $S S_3$ (Model II)
in the superpotential. These again would not affect our conclusions
if present, and in any case bilinear terms  are not expected in 
(conformal) string theory.}.

Further details of the Higgs masses and eigenstates are given in the Appendices.

\section{Specific Models}
\label{models}
\subsection{Model I}
In Model I, we choose the $U(1)'$ charges for the Higgs fields as
\begin{eqnarray}
Q_S=-Q_{S_1} =-Q_{S_2} ={1\over 2} Q_{S_3} ~,~ Q_{H_1}+Q_{H_2}+Q_S=0
~.~\,
\end{eqnarray} 
The 
 dimension-2 supersymmetry breaking soft terms
\begin{eqnarray}
V_{soft}^{(I)} &=& (m_{S S_1}^2 S S_1 + m_{S S_2}^2 S S_2 
+ m_{S_1 S_2}^2 S_1^{\dagger} S_2 + {\rm H. C.})
~\,
\end{eqnarray}
are allowed by the \upr, so in general
\begin{eqnarray}
V_{soft} &=& V_{soft}^{(a)} + V_{soft}^{(I)}
~,~\,
\end{eqnarray} 
where $V_{soft}^{(a)}$ is defined in (\ref{vsoft}).
 However, only two of these are needed 
 to break the  global $U(1)$ symmetries, so for simplicity we
will set\footnote{Keeping a nonzero $m_{S_1 S_2}^2$ would yield a
spectrum similar to Model II. In the most general case  one
would have to allow $m_{S_1 S_2}^2$ to be complex valued
(and therefore CP violating) because
there would not be
enough freedom of field redefinitions to 
ensure that all three terms are real and negative. }
$m_{S_1 S_2}^2=0$.

To avoid directions of the potential that are not bounded from
below, we
require
\begin{eqnarray}
m_S^2 + m_{S_1}^2 +2 m_{S S_1}^2 > 0 ~,~ m_S^2 + m_{S_2}^2 +2 m_{S S_2}^2 > 0
~.~\,
\end{eqnarray} 
The first condition corresponds to the direction in which $s = s_1$ with
the other VEVs vanishing, for which the quartic and cubic terms in the
potential are flat. The second corresponds to $s = s_2$.

The minimization conditions for the neutral scalar potential with non-zero
VEVs, and the mass matrices for the CP odd and even Higgses are given in
Appendix~A.

\subsection{Model II}
In Model II, we choose the $U(1)'$ charges
\begin{eqnarray}
Q_{S_1} = Q_{S_2} =   {1\over 2} Q_{S} = - {1\over 2} Q_{S_3} ~,~ Q_{H_1}+Q_{H_2}+Q_S=0
~,~\,
\end{eqnarray} 
allowing the dimension-2 supersymmetry terms
\begin{eqnarray}
V_{soft}^{(II)} &=& (m_{S S_3}^2 S S_3 + 
m_{S_1 S_2}^2 S_1^{\dagger} S_2 + {\rm H. C.})
~,~\,
\end{eqnarray} 
so that
\begin{eqnarray}
V_{soft} &=& V_{soft}^{(a)} + V_{soft}^{(II)}
~.~\,
\end{eqnarray} 

To avoid unbounded from below directions, we require
\begin{eqnarray}
m_S^2 + 2 m_{S_1}^2 > 0 ~,~ m_S^2 + 2 m_{S_2}^2 > 0
~,~\,
\end{eqnarray} 
corresponding to the directions with $s_1 = \sqrt{2}s$ and
$s_2 = \sqrt{2}s$ (and the other VEVs zero), respectively.

The minimization conditions and  mass matrices
are similar to those in Model~I up to obvious changes, 
so they will not be repeated.

\subsection{Numerical Results for Some Particle Spectra}

\renewcommand{\arraystretch}{1.4}
\begin{table}[t]
\caption{$v_1$, $v_2$, $s$, $s_1$, $s_2$, $s_3$; the  $Z$ and $Z'$ masses; and
$\alpha_{Z-Z'}$ in  Models I and II. The masses and VEVs  are in GeV. 
\label{tab:mix}}
\vspace{0.4cm}
\begin{center}
\begin{tabular}{|c|c|c|c|c|c|c|c|c|c|c|}
\hline        
 Model & $h$ & $v_1$ & $v_2$ & $s$ & $s_1$ & $s_2$ & $s_3$
 & $Z$ & $Z^{\prime}$ & $\alpha_{Z-Z'}$ \\
\hline
I & 0.5 & 121 & 125 & 187  & 1270 & 1260 & 1260 & 91  & 2030  & $3.8\times
10^{-3}$ \\
\hline
I & 0.75 & 121  & 125 & 187  & 1270 & 1260 & 1260  & 91  & 2030  
& $3.8\times 10^{-3}$ \\
\hline
II & 0.5 & 122  & 124 & 175  & 1300  & 1300 & 1290 & 91  & 2100  
& $4.7\times 10^{-3}$ \\
\hline
II & 0.75 & 122  & 124  & 178  & 1310  & 1310 & 1300 & 91  & 2110  
& $4.7\times 10^{-3}$ \\
\hline
\end{tabular}
\end{center}
\end{table}

\renewcommand{\arraystretch}{1.4}
\begin{table}[t]
\caption{The chargino and neutralino masses in  GeV for Models I and II. 
\label{tab:neut}}
\vspace{0.4cm}
\begin{center}
\begin{tabular}{|c|c|c|c|c|c|c|c|c|c|c|c|c|c|}
\hline        
Model & $h$ & $M_i$ & $\tilde \chi_1^{\pm}$ & $\tilde \chi_2^{\pm}$ 
 & $\tilde \chi_1^{0}$  & $\tilde \chi_2^{0}$ & $\tilde \chi_3^{0}$  &
  $\tilde \chi_4^{0}$ & $\tilde \chi_5^{0}$ &  $\tilde \chi_6^{0}$ &
 $\tilde \chi_7^{0}$ & $\tilde \chi_8^{0}$ & $\tilde \chi_9^{0}$ \\
\hline
 I & 0.5 & $<$ 0 &114 & 220 & 52 & 63 & 107 & 122 & 126 & 145 & 221 & 1790 & 2320
\\
\hline
 I & 0.5 & $>$ 0 & 74 & 420 & 52 & 61 & 63 & 126 & 145 & 213 & 420 & 1710 & 2380\\
\hline
 I & 0.75 & $<$ 0 & 158 & 218 & 78 & 94 & 106 & 165 & 189 & 218 & 219 & 1800 & 2310
\\
\hline
 I & 0.75 & $>$ 0 & 118 & 423 & 78 & 94 & 100 & 189 & 217 & 218 & 423 & 1700 & 2390
\\
\hline
 II & 0.5 & $<$ 0 & 108 & 221 & 54 & 65 & 107 & 116 & 130 & 141 & 222 & 1860  &
2390 \\
\hline
 II & 0.5 & $>$ 0 & 68  & 419 & 54 & 56 & 65 & 130 & 141 & 212 & 420 & 1780 & 2450
\\
\hline
 II & 0.75 & $<$ 0 & 152  & 218 & 80 & 98 & 106 & 158 & 196 & 213 & 219 & 1890 &
2390 \\
\hline
 II & 0.75 & $>$ 0 & 111  & 422 & 80 & 94 & 98 & 196 & 213 & 216 & 423 & 1780 &
2480 \\
\hline
\end{tabular}
\end{center}
\end{table}

\renewcommand{\arraystretch}{1.4}
\begin{table}[t]
\caption{The charged, CP even, and  CP odd
 Higgs masses in GeV at tree-level  for  Models I and II.
\label{tab:Higgs}}
\vspace{0.4cm}
\begin{center}
\begin{tabular}{|c|c|c|c|c|c|c|c|c|c|c|c|c|}
\hline        
Model & $h$ & $H^{\pm}$ & $H_1^0$  & $H_2^0$  & $H_3^0$  & $H_4^0$ 
  & $H_5^0$  & $H_6^0$ & $A_1^0$ & $A_2^0$ & $A_3^0$ & $A_4^0$  \\
\hline
I& 0.5 & 152 & 52 & 88 & 92 & 112  & 158  & 2030 & 5.0 & 43 & 157 & 174 \\     
\hline
I& 0.75 & 211 & 78 & 131 & 139 & 168 & 215 & 2030 & 7.6 & 65 & 236 & 261 \\     
\hline
II& 0.5 & 146 & 59 & 92 & 93 & 108 & 152 & 2100 & 22 & 38 & 158 & 168 \\     
\hline
II& 0.75 & 203 & 86 & 139 & 140 & 162 & 207 & 2120 & 34 & 58 & 239 & 255  \\ 
\hline
\end{tabular}
\end{center}
\end{table}

In this subsection, we  present the numerical results for the 
$Z'$ boson mass; the $Z-Z'$ mixing angle $\alpha_{Z-Z'}$; and the chargino,
neutralino, and Higgs masses for the two models. 
To generate the mass hierarchy between the
$Z$ and $Z'$,
we choose  $\lambda = h/10$. We illustrate for two values of $h$, i.e.,
0.5 and 0.75. Both are theoretically and phenomenologically  acceptable.
However, the larger value allows larger chargino and neutralino
masses, but is close to the upper limit allowed if the theory
is to remain perturbative to a large scale.
In our conventions, $\mu > 0$, while the gaugino masses $M_i$ can
be positive or negative. We choose two examples, i.e.,
$M_1=-100$ GeV, $M_2=-200$ GeV and $M_1^{\prime}=-600$ GeV; and
$M_1=200$ GeV, $M_2=400$ GeV and $M_1^{\prime}=600$ GeV.
These choices  can yield relatively large
masses for the lighter charginos.
We also choose the standard GUT
value  $g_{Z'}=\sqrt{5/3} g_1$ (it is
$\sqrt{5/3} g_1$ that unifies with $g_2$ and $g_3$ in the simple
GUT models). This is for illustration only; we do not insist
on conventional grand unification\footnote{Many unification models would
suggest $M_1^{\prime}\sim M_1$. The only effect of a smaller $M_1^{\prime}$
would be small changes in the spectrum of the neutralinos in the secluded sector.}.
As described above, we choose
large values for $A_h$ (and also choose $A_\lambda \sim A_h$).
Otherwise, the  minimum of the potential would be for $v_1 = v_2 = 0$ and
$s \sim s_i$. (Even for large $A_h$ there is such a local minimum.
However, there is also the desired $SU(2)$-breaking minimum closer
to the origin.) The terms linear in $s$ in
$V_{soft}^{(I,II)}$ and the $A_h$ term
prevent unphysical minima such as $s=v_1=0$, $s=v_2=0$, or only one of the
three vanishing.

The input parameters with dimensions or mass of mass-squared are chosen
in arbitrary units. After finding an acceptable minimum they are rescaled
so that $\sqrt{v_1^2+v_2^2}\simeq 174$ GeV.
For Model I, we choose:
$A_h = A_{\lambda} =1.0,
m_{H_1}^2 = m_{H_2}^2= m_{S}^2 = -0.010,
m_{S_1}^2 = m_{S_2}^2=0.031, m_{S_3}^2 = -0.010,
m_{S S_1}^2= m_{S S_2}^2=-0.010, Q_{H_1}=1,
Q_{H_2}=-2, Q_S=- Q_{S_1}=- Q_{S_2}=1$,  and $Q_{S_3}=2$.

For $h=0.5$ and $\lambda=0.05$
the VEVs at the  minimum are 
$v_1 =0.928, v_2=0.953,
s=1.43, s_1=9.67, s_2=9.65$, and $s_3=9.63$.
For $h=0.75$ and $\lambda=0.075$ they are $v_1 =0.616, v_2=0.636,
s=0.953, s_1=6.44, s_2=6.42$, and $s_3=6.41$.
The rescaled VEVs and the corresponding \zpr \ mass and $Z-\zpr$
mixing angle are listed in Table~\ref{tab:mix}.

For  Model II, we choose
$ A_h = A_{\lambda} =1.0,
m_{H_1}^2 = m_{H_2}^2=-0.010, m_{S}^2 = -0.020,
m_{S_1}^2 = m_{S_2}^2=0.011, m_{S_3}^2 = -0.010,
m_{S S_3}^2= m_{S_1 S_2}^2=-0.015, Q_{H_1}=1,
Q_{H_2}=-3, Q_S=2, Q_{S_1}= Q_{S_2}= 1, Q_{S_3}=-2$.
The VEVs are
$v_1 =0.955, v_2=0.965,
s=1.37, s_1=10.1, s_2=10.1, s_3=10.1$
for  $h=0.5$ and $\lambda=0.05$;
and $v_1 =0.632, v_2=0.638,
s=0.919, s_1=6.76, s_2=6.75, s_3=6.73$
 for $h=0.75$ and $\lambda=0.075$.


The rescaled  VEVs $v_1$, $v_2$, $s$, $s_1$, $s_2$, $s_3$; the mixing $\alpha_{Z-Z'}$; and
the particle spectra are given in Tables~\ref{tab:mix},~\ref{tab:neut},
and~\ref{tab:Higgs}.
It is seen that the two models yield similar spectra, since each is
$A_h$ dominated.
The composition of the physical mass eigenstates in terms of the
weak eigenstates are given for Model~I in Appendix B.
The spectra are quite different from the MSSM. The most important feature is that
the VEVs of the secluded sector fields ($S_1, S_2$, and $S_3$)
 are much larger than those of the ordinary sector ($H_i$ and
$S$), without any fine tuning of parameters. This leads to a rather heavy \zpr, a
small $\alpha_{Z-Z'}$ ($\alpha_{z-z'}$ would have been zero or extremely
small,
depending on the soft mass-squares,
if we had chosen  $Q_{H_1} = Q_{H_2}$), and little mixing between the ordinary and secluded
sectors. The large $A_h$ needed to ensure the correct vacuum implies $\tan \beta \sim 1$
and $s/v_i \lsim 3/2$, leading to significant mixing between the doublet and
singlet Higgs fields, and also between the corresponding neutralinos. 

The upper limit on the lightest CP even Higgs doublet particle is considerably relaxed
compared to the MSSM and even the NMSSM. However, the actual mass eigenvalues are
reduced by mixing with the $SU(2)$ singlet. For example, in Model I with $h = 0.5$,
the lightest scalar, $H_1^0$,  and $H_4^0$ are roughly equal admixtures of singlet
($S$) and doublets; $H_5^0$ is almost pure doublet; while the two light
states $H_{2,3}^0$ and the very heavy $H_6^0$ consist almost entirely of the
secluded fields. Similarly, the CP odd states $A_2^0$ and $A_4^0$ are 
ordinary-sector doublet-singlet mixtures, while the very light $A_1^0$ and
the heavier $A_3^0$ consist mainly of the $S_i$. (The small values
for the $A_1^0$ mass reflect the fact that the  off-diagonal scalar
masses added to
break the two global symmetries were chosen to be small compared
to $A_h$ and $A_\lambda$, and, in the case of Model I, that the terms involve 
$s \ll s_{1,2}$.) 
The lightest CP even Higgs has tree-level mass $\sim$ 52 GeV for Model I with $h = 0.5$.
It is not clear whether this is consistent with experiment. In the first place, the
masses may increase significantly due to radiative corrections analogous to
those of the MSSM. However, the actual values depend on parameters associated
with the sfermion sector of the model, which we are not considering here. Also, 
the state involves a large admixture of $SU(2)$ singlet, so the usual 
SM and MSSM limits do not apply directly, and will require a detailed
study of the collider implications of such mixings that is beyond the scope
of this paper. Of course, the mass could be increased somewhat for different
choices of the soft parameters.

The chargino masses are consistent with the experimental limits except for
the cases with $h=0.5$ and $M_2 > 0$. The lighter chargino is dominantly
Higgsino for our choices of $M_2$, because of the relatively low effective $\mu$ parameter.

In the first row of Table~\ref{tab:neut}, the lightest neutralino
$\tilde{\chi}_1^0$ 
is mainly singlino $\tilde{S}$, with a nontrivial
admixture of Higgsino. This is somewhat similar to the model in~\cite{dce}, in which
the light singlino was advocated as a dark matter candidate.
$\tilde{\chi}_6^0$ is also a mixture, while
$\tilde{\chi}_4^0$, $\tilde{\chi}_3^0$, and $\tilde{\chi}_7^0$ 
are mainly Higgsino, bino, and wino, respectively. The actual composition of these
ordinary sector states is affected by the low effective $\mu$, but  also depends
significantly on the choice of gaugino mass inputs. 
$\tilde{\chi}_2^0$  and $\tilde{\chi}_5^0$ are mainly secluded sector
singlinos, while the two heavy states 
$\tilde{\chi}_8^0$  and $\tilde{\chi}_9^0$ are admixtures of $\tilde{B}^\prime$
and singlino ($\tilde{S}_i$).

The soft masses for squarks and sleptons are independent of the
symmetry breaking sector. However, if they are chosen to be of 
the same order as $A_h$ (i.e., larger than the soft masses of the doublet and singlet Higgs 
fields) then they would typically be in the 100--300 GeV range.

Clearly, the spectrum of the symmetry breaking sector is very rich, and will differ
significantly from that of the MSSM because of the significant doublet-singlet mixing.
A detailed study of the collider signatures and limits  and the implications for
cold dark matter is beyond the scope of this paper. However, 
there will generically be a number of particles associated
with the symmetry breaking sector that are on the margin of being excluded or
discovered. A  further study would be very interesting.

\section{Discussion and Conclusion} 
\label{discussion}

Many theories beyond the Standard Model predict the existence of addition
\upr \ gauge symmetries broken near the electroweak scale. 
Although such models have the desirable feature of  yielding a simple
solution to the $\mu$ problem, they suffer from the need to make the
\zpr \ sufficiently heavy,  typically at least (500-800) GeV.
Previous models have
often assumed that this is accomplished  by  having a typical soft supersymmetry
breaking scale (and corresponding sfermion masses) of a TeV or so, with the
electroweak scale smaller by cancellations. In this paper, we present
a different mechanism, in which the large \upr \ breaking scale is associated
with an almost flat direction of the quartic terms of the scalar potential.
In the limit that a certain Yukawa coupling  goes to zero, this
could correspond to a runaway, unbounded
from below direction. The flatness is lifted by small but nonzero values, allowing
a large $\zpr/Z$ mass hierarchy and a small 
(or almost zero for some charge assignments)
$Z-Z^\prime$ mixing angle.

We have presented examples of such models involving
 an ordinary sector of symmetry breaking fields,
which includes two Higgs doublets and an $SU(2)$
singlet $S$ which generates an effective $\mu$
parameter; and a secluded sector involving three $SU(2)$ singlet fields
$S_i, i = 1,2,3$,
which acquire large VEVs. The two sectors are only weakly coupled by \upr \
interactions
and soft scalar mass terms. The ordinary sector is somewhat similar to the
NMSSM, but involves parameter choices very different from those usually
studied in the NMSSM context. (These include the absence of a cubic
term in $S$ in the superpotential, a larger allowed value for $h$, and 
a large $A_h$.)
We carried out a detailed study of such issues as unwanted
global minima and runaway directions,
unwanted global symmetries, the upper limit from
perturbative unification on
the Yukawa coupling
 associated with the effective $\mu$ parameter, and the need to have sufficiently
heavy charginos.  Acceptable parameter ranges were found, characterized by:
 the electroweak
symmetry breaking is driven
more by a large $A$ term than by the soft scalar mass-squares,
leading to $\tan \beta \sim 1$; a VEV of $S$ 
comparable to the electroweak scale;
a fairly small effective $\mu$ parameter (typically 80--140 GeV);
and a much larger \upr \ breaking scale generated by the VEVs of the $S_i$.

The spectrum of the symmetry breaking sector is very rich. There are a number of
light CP even and odd Higgs fields and neutralinos, for example, which involve
significant
mixing between $SU(2)$ doublet and singlet fields. A detailed study of the implications
for colliders and cosmology is beyond the scope of this paper, but it is expected
that a number of the predicted states are close to being excluded or discovered.

We have also not attempted to embed the models into a full theory. This would be
necessary to discuss the sfermion spectrum; the cancellation of anomalies;
possible flavor changing effects~\cite{mp}; and
some aspects of the production and decay of the Higgs particles, charginos, and
neutralinos.

\section*{Acknowledgments}
This research was supported in part by the U.S.~Department of Energy under
 Grants No.~DOE-EY-76-02-3071 and DE-FG02-95ER40896, and by the University of
Wisconsin at Madison.

\section*{Appendix A}
In Appendix A we discuss the 
minimization conditions and scalar mass matrices for Model I.
The conditions for Model II are similar.

The potential minimization conditions for the neutral scalar fields with non-zero
VEVs are
\begin{eqnarray}
m_{H_1}^2 - A_h h s v_2/v_1 +h^2 (v_2^2 +s^2) +{{G^2}\over 4} (v_1^2-v_2^2)
+g_{Z'}^2 Q_{H_1} \Delta =0
~,~\,
\end{eqnarray} 
\begin{eqnarray}
m_{H_2}^2 - A_h h s v_1/v_2 +h^2 (v_1^2 +s^2) +{{G^2}\over 4} (v_2^2-v_1^2)
+g_{Z'}^2 Q_{H_2} \Delta =0
~,~\,
\end{eqnarray} 
\begin{eqnarray}
m_{S}^2 - A_h h v_1 v_2/s + m_{S S_1}^2 s_1/s + m_{S S_2}^2 s_2/s 
+ h^2 (v_1^2 +v_2^2) +g_{Z'}^2 Q_{S} \Delta =0
~,~\,
\end{eqnarray} 
\begin{eqnarray}
m_{S_1}^2 - A_{\lambda} \lambda s_2 s_3/s_1 + m_{S S_1}^2 s/s_1 
+ \lambda^2 (s_2^2 +s_3^2) + g_{Z'}^2 Q_{S_1} \Delta =0
~,~\,
\end{eqnarray} 
\begin{eqnarray}
m_{S_2}^2 - A_{\lambda} \lambda s_1 s_3/s_2 + m_{S S_2}^2 s/s_2
+ \lambda^2 (s_1^2 +s_3^2) + g_{Z'}^2 Q_{S_2} \Delta =0
~,~\,
\end{eqnarray}
\begin{eqnarray}
m_{S_3}^2 - A_{\lambda} \lambda s_1 s_2/s_3  
+ \lambda^2 (s_1^2 +s_2^2) + g_{Z'}^2 Q_{S_3} \Delta =0
~.~\,
\end{eqnarray}

The mass-square matrix for the CP odd neutral Higgs particles
in the basis $\{H_1^{0i} \equiv \sqrt 2 Im(H_1^0),
H_2^{0i}, S^{0i}, S_1^{0i}, S_2^{0i}, S_3^{0i}\}$ is
\begin{eqnarray}
M_{A^{0}}^2 =\left(\matrix{ O_{A^{0}}
& C_{A^{0}} \cr
C_{A^{0}}^T& S_{A^{0}} \cr}\right)
~,~ \,
\end{eqnarray}
where
\begin{eqnarray}
O_{A^{0}} = \left(\matrix{A_h h s v_2/v_1 & A_h h s & A_h h v_2 \cr
A_h h s & A_h h s v_1/v_2 & A_h h v_1 \cr
A_h h v_2 & A_h h v_1 & \beta_{S}^2 \cr }\right) ,\,
\end{eqnarray}
\begin{eqnarray}
S_{A^{0}}= \left(\matrix{
\beta_{S_1}^2 & 
A_{\lambda} \lambda s_3 & A_{\lambda} \lambda s_2 \cr
A_{\lambda} \lambda s_3 & 
\beta_{S_2}^2& 
A_{\lambda} \lambda s_1 \cr
A_{\lambda} \lambda s_2 & 
A_{\lambda} \lambda s_1 & 
A_{\lambda} \lambda s_1 s_2/s_3 \cr}\right) ,\,
\end{eqnarray}
\begin{eqnarray}
C_{A^{0}}= \left(\matrix{0 & 0& 0\cr
0 & 0 & 0\cr
-m^2_{S S_1} & -m^2_{S S_2} & 0 \cr }\right) ,\,
\end{eqnarray}
and
\begin{eqnarray}
\beta_{S}^2 = (A_h h v_1 v_2-m^2_{S S_1} s_1 -
m^2_{S S_2} s_2)/s
~,~\,
\end{eqnarray} 
\begin{eqnarray}
\beta_{S_1}^2 = A_{\lambda} \lambda s_2 s_3/s_1-m^2_{S S_1} s/s_1
~,~\,
\end{eqnarray} 
\begin{eqnarray}
\beta_{S_2}^2 = A_{\lambda} \lambda s_1 s_3/s_2-m^2_{S S_2} s/s_2
~.~\,
\end{eqnarray}

Similarly, in the basis
$\{H_1^{0r} \equiv \sqrt 2 Re(H_1^0),
H_2^{0r}, S^{0r}, S_1^{0r}, S_2^{0r}, S_3^{0r}\}$,  the
mass-square matrix for the CP even neutral Higgs particles is
\begin{eqnarray}
M_{H^{0}}^2 =\left(\matrix{ O_{H^{0}}
& C_{H^{0}} \cr
C_{H^{0}}^T& S_{H^{0}} \cr}\right)
~,~ \,
\end{eqnarray}
where
\begin{eqnarray}
O_{H^{0}}= \left(\matrix{
\kappa_{H_1}^2  & \kappa_{H_1, H_2}  
& \kappa_{H_1, S} \cr
\kappa_{H_1, H_2} & \kappa_{H_2}^2  &
\kappa_{H_2, S}  \cr
\kappa_{H_1, S}  & 
\kappa_{H_2, S}  & \kappa_S^2  \cr }\right) ,\,
\end{eqnarray}
\begin{eqnarray}
S_{H^{0}}= \left(\matrix{
\kappa_{S_1}^2 & 
\kappa_{S_1, S_2}& 
\kappa_{S_1, S_3}  \cr
\kappa_{S_1, S_2} & 
\kappa_{S_2}^2 & 
\kappa_{S_2, S_3}\cr
\kappa_{S_1, S_3}  & 
\kappa_{S_2, S_3} & 
\kappa_{S_3}^2 \cr}\right) ,\,
\end{eqnarray}
\begin{eqnarray}
C_{H^{0}}= \left(\matrix{
\kappa_{H_1, S_1} & \kappa_{H_1, S_2} & \kappa_{H_1, S_3} \cr
\kappa_{H_2, S_1} & \kappa_{H_2, S_2} & \kappa_{H_2, S_3} \cr 
\kappa_{S, S_1} +m^2_{S, S_1} & 
\kappa_{S, S_2}+m^2_{S, S_2} & \kappa_{S, S_3} \cr}\right) ,\,
\end{eqnarray}
and
\begin{eqnarray}
\kappa_{H_i}^2 = 2 \left( {{G^2} \over {4}} + g_{Z'}^2 Q_{H_i}^2 
\right) v_i^2 + A_h h s v_1 v_2/v_i^2
~,~\,
\end{eqnarray} 
\begin{eqnarray}
\kappa_S^2 = 2 g_{Z'}^2 Q_S^2 s^2 + (A_h h v_1 v_2-m^2_{S S_1} s_1 - 
m^2_{S S_2} s_2)/s
~,~\,
\end{eqnarray} 
\begin{eqnarray}
\kappa_{S_1}^2 = 2 g_{Z'}^2 Q_{S_1}^2 s_1^2 +
 A_{\lambda} \lambda s_2 s_3/s_1-m^2_{S S_1} s/s_1
~,~\,
\end{eqnarray} 
\begin{eqnarray}
\kappa_{S_2}^2 = 2 g_{Z'}^2 Q_{S_2}^2 s_2^2 +
 A_{\lambda} \lambda s_1 s_3/s_2-m^2_{S S_2} s/s_2
~,~\,
\end{eqnarray} 
\begin{eqnarray}
\kappa_{S_3}^2 = 2 g_{Z'}^2 Q_{S_3}^2 s_3^2 + A_{\lambda} \lambda s_1 s_2/s_3
~,~\,
\end{eqnarray} 
\begin{eqnarray}
\kappa_{H_1, H_2} = 2 \left( h^2 - {{G^2} \over {4}} + 
g_{Z'}^2 Q_{H_1} Q_{H_2} \right) v_1 v_2 - A_h h s 
~,~\,
\end{eqnarray} 
\begin{eqnarray}
\kappa_{H_i, S} = 2 \left( h^2 + 
g_{Z'}^2 Q_{H_i} Q_S \right) v_i s - |\epsilon_{i j}| A_h h v_j
~,~\,
\end{eqnarray}
\begin{eqnarray}
\kappa_{H_i, S_j} = 2 g_{Z'}^2 Q_{H_i} Q_{S_j} v_i s_j ~,~ 
\kappa_{S, S_i} = 2 g_{Z'}^2 Q_S Q_{S_i} s s_i
~,~\,
\end{eqnarray} 
\begin{eqnarray}
\kappa_{S_i, S_j} = 2 (\lambda^2 + g_{Z'}^2 Q_{S_i} Q_{S_j}) s_i s_j
-|\epsilon_{ijk}| A_{\lambda} \lambda s_k 
~.~\,
\end{eqnarray} 
The upper limit in (\ref{doubletlimit}) on the lightest doublet Higgs mass is obtained from the
limit on the smaller eigenvalue of the upper $2 \times 2$ sub-block of $O_{H^{0}}$.

\section*{Appendix B}
   
The chargino mass terms are~\cite{BARTL}
\begin{eqnarray}
\cal L &=& - (\psi^-)^T M_{\tilde \chi^{\pm}} \psi^+ + {\rm H.C.} ~,~ \,
\end{eqnarray}
where $(\psi^{+})^T=(-i \tilde W^{+}, \tilde H^{+}_2)$ and
$(\psi^{-})^T=(-i \tilde W^{-}, \tilde H^{-}_1)$
are two component spinors, and $M_{\tilde \chi^{\pm}}$ is given
in Eq. (16). The chargino mass eigenstates are defined by
\begin{eqnarray}
\tilde \chi_i^+ = V_{ij} \psi_j^+ ~;~
 \tilde \chi_i^- = U_{ij} \psi_j^- ~,~ \,
\end{eqnarray} 
 where $U$ and $V$ are unitary matrices.

The eigenvectors for the charginos; neutralinos; 
and CP even and  odd Higgses in Model I
with $h=0.5$ and $h=0.75$ are 
given in Tables \ref{tab:ech}-\ref{tab:eA2}.

\renewcommand{\arraystretch}{1.4}
\begin{table}[t]
\caption{The eigenvectors for the  charginos in Model I.
\label{tab:ech}}
\vspace{0.4cm}
\begin{center}
\begin{tabular}{|c|c|c|c|c|c|c|c|c|c|}
\hline        
 $h$ & $M_i$ & $U_{11}$ & $U_{12}$ & $U_{21}$ & $U_{22}$
& $V_{11}$ & $V_{12}$ & $V_{21}$ & $V_{22}$ \\
\hline
  0.5 & $<$ 0 & 0.237 & 0.971 & 0.971 & -0.237 & 0.257 & 0.967
& -0.967 & 0.257\\
\hline
  0.5 & $>$ 0 & -0.236 & 0.972 & 0.972 & 0.236 & 0.240 & -0.971
& 0.971 & 0.240 \\
\hline
 0.75 & $<$ 0 & 0.197 & 0.980 & 0.980 & -0.197 & 0.238 & 0.971
& -0.971 & 0.238 \\
\hline
 0.75 & $>$ 0 & -0.270 & 0.963 & 0.963 & 0.270 & 0.275 & -0.962
& 0.962 & 0.275\\
\hline
\end{tabular}
\end{center}
\end{table}

\renewcommand{\arraystretch}{1.4}
\begin{table}[t]
\caption{ The eigenvectors for the neutralinos in Model I
with $h=0.5$ and $M_i < 0$.
\label{tab:en1}}
\vspace{0.4cm}
\begin{center}
\begin{tabular}{|c|c|c|c|c|c|c|c|c|c|}
\hline        
Fields  & $\tilde \chi_1^{0}$  & $\tilde \chi_2^{0}$ & $\tilde \chi_3^{0}$  &
  $\tilde \chi_4^{0}$ & $\tilde \chi_5^{0}$ &  $\tilde \chi_6^{0}$ &
 $\tilde \chi_7^{0}$ & $\tilde \chi_8^{0}$ & $\tilde \chi_9^{0}$ \\
\hline
${\tilde B}^{\prime}$ & 0.0 & 0.0 & 0.001 & 0.002 & 0.0 &
0.001 & -0.003 & -0.647 & 0.762\\
\hline
$\tilde B$ & -0.004 & 0.0 & 0.978 & 0.188 & 0.0 &
-0.007 & -0.093  & 0.002 & -0.001\\
\hline
$\tilde W_3^0$ & 0.005 & 0.0 & 0.137 & -0.237 & 0.0 &
-0.010 &  0.962 & -0.003 & 0.002\\
\hline
$\tilde H_1^0$ & -0.356 & 0.0 & 0.116 & -0.675 & 0.0 &
0.610 & -0.174  & -0.030 & -0.025\\
\hline
$\tilde H_2^0$ & -0.368 & 0.0 & -0.108 & 0.668 & 0.0 &
0.604 & 0.188 & 0.062 & 0.051\\
\hline
$\tilde S$ & 0.856 & 0.001 & 0.005 & 0.015 & 0.0 &
0.513 & 0.004  & -0.045 & -0.039\\
\hline
$\tilde S_1$ & 0.028 & -0.707 & 0.005 & -0.033 & 0.577 &
0.003 & -0.009 & 0.310 & 0.264\\
\hline
$ \tilde S_2$ & 0.027 & 0.707 & 0.005 & -0.033 & 0.577 &
0.003 & -0.009 & 0.309 & 0.263\\
\hline
$\tilde S_3$ & -0.055 & -0.001 & -0.011 & 0.066 & 0.578 &
-0.005 & 0.017 & -0.618 & -0.525\\
\hline
\end{tabular}
\end{center}
\end{table}

\renewcommand{\arraystretch}{1.4}
\begin{table}[t]
\caption{ Same as Table~\ref{tab:en1}, except
$M_i > 0$.
\label{tab:en2}}
\vspace{0.4cm}
\begin{center}
\begin{tabular}{|c|c|c|c|c|c|c|c|c|c|}
\hline        
Fields  & $\tilde \chi_1^{0}$  & $\tilde \chi_2^{0}$ & $\tilde \chi_3^{0}$  &
  $\tilde \chi_4^{0}$ & $\tilde \chi_5^{0}$ &  $\tilde \chi_6^{0}$ &
 $\tilde \chi_7^{0}$ & $\tilde \chi_8^{0}$ & $\tilde \chi_9^{0}$ \\
\hline
${\tilde B}^{\prime}$ & 0.0 & 0.0 & 0.0 & 0.0 & 0.001 &
 0.002 & -0.004 & 0.659 & 0.752\\
\hline
$\tilde B$ & 0.036 & 0.290 & 0.001 & 0.0 & -0.001 & 
0.955 & -0.048 & -0.001 & -0.001\\
\hline
$\tilde W_3^0$ & -0.028 & -0.217 & -0.001 & 0.0 & 0.001 &
0.116 & 0.969 & 0.002 & 0.002\\
\hline
$\tilde H_1^0$ & -0.279 & 0.697 & 0.002 & 0.0 & 0.608 & 
-0.192 & 0.170 & -0.029 & 0.026\\
\hline
$\tilde H_2^0$ & -0.440 & -0.605 & -0.002 & 0.0 & 0.606 & 
0.192 & -0.172 & 0.059 & -0.054\\
\hline
$\tilde S$ & 0.849 & -0.111 & 0.0 & 0.0 & 0.513 & 
0.002 & -0.001 & -0.046 & 0.039\\
\hline
$\tilde S_1$ & 0.032 & 0.031 & -0.707 & 0.577 & 0.003 &
-0.009 & 0.008 & 0.306 & -0.268\\
\hline
$\tilde S_2$ & 0.031 & 0.026 & 0.707 & 0.577 & 0.003 &
-0.009 & 0.008 & 0.305 & -0.268\\
\hline
$\tilde S_3$ & -0.062 & -0.057 & -0.001 & 0.578 & -0.005 &
0.018 & -0.016 & -0.610 & 0.535\\
\hline
\end{tabular}
\end{center}
\end{table}

\renewcommand{\arraystretch}{1.4}
\begin{table}[t]
\caption{ Same as Table~\ref{tab:en1}, except
$h=0.75$.
\label{tab:en3}}
\vspace{0.4cm}
\begin{center}
\begin{tabular}{|c|c|c|c|c|c|c|c|c|c|}
\hline        
Fields  & $\tilde \chi_1^{0}$  & $\tilde \chi_2^{0}$ & $\tilde \chi_3^{0}$  &
  $\tilde \chi_4^{0}$ & $\tilde \chi_5^{0}$ &  $\tilde \chi_6^{0}$ &
 $\tilde \chi_7^{0}$ & $\tilde \chi_8^{0}$ & $\tilde \chi_9^{0}$ \\
\hline
${\tilde B}^{\prime}$ & 0.001 & 0.0 & 0.001 & -0.002 & 0.0 &
 -0.003 & -0.003 & 0.662 & 0.750\\
\hline
$\tilde B$ & -0.005  & 0.0 & 0.984 & -0.065 &  0.0 &   
 -0.052 & -0.159 & -0.002 & -0.001\\
\hline
$\tilde W_3^0$ & 0.006  & 0.0 & 0.117 & 0.734 & 0.0 &        
0.635 & 0.211 &  0.003 & 0.002\\
\hline
$\tilde H_1^0$ & -0.352 & 0.0 & 0.098 & 0.279 & 0.0 &          
-0.564 & 0.685 & -0.029 & 0.026\\
\hline
$\tilde H_2^0$ & -0.372 & 0.0 & -0.096 & 0.517 & 0.0 &          
-0.353 & -0.674 & 0.059 & -0.054\\
\hline
$\tilde S$ &   0.856 & 0.001 & 0.003 & 0.334 &  0.0 &      
-0.389 &  -0.020 & -0.046 & 0.039\\
\hline
$\tilde S_1$ & 0.029 & -0.707 & 0.005 & -0.005 & 0.576 &      
-0.008 & 0.032 & 0.305 & -0.270\\
\hline
$\tilde S_2$ & 0.027 & 0.707 & 0.005 & -0.005 &  0.577 &         
-0.008 & 0.033 & 0.304 & -0.269\\
\hline
$\tilde S_3$ & -0.056 & -0.001 & -0.010 & 0.009 &  0.578 &      
0.015 & -0.065 & -0.608 & 0.537\\
\hline
\end{tabular}
\end{center}
\end{table}

\renewcommand{\arraystretch}{1.4}
\begin{table}[t]
\caption{ Same as Table~\ref{tab:en1}, except
$h=0.75$ and $M_i > 0$.
\label{tab:en4}}
\vspace{0.4cm}
\begin{center}
\begin{tabular}{|c|c|c|c|c|c|c|c|c|c|}
\hline        
Fields  & $\tilde \chi_1^{0}$  & $\tilde \chi_2^{0}$ & $\tilde \chi_3^{0}$  &
  $\tilde \chi_4^{0}$ & $\tilde \chi_5^{0}$ &  $\tilde \chi_6^{0}$ &
 $\tilde \chi_7^{0}$ & $\tilde \chi_8^{0}$ & $\tilde \chi_9^{0}$ \\
\hline
${\tilde B}^{\prime}$ & 0.001 & 0.0 & 0.0 & 0.0 & 0.002 &
 0.001 & -0.004 & 0.662 & 0.750\\
\hline
$\tilde B$ & 0.029 & -0.001 & 0.388 & 0.0 & 0.920 & 
 -0.001 & -0.055 & -0.001 & -0.001\\
\hline
$\tilde W_3^0$ & -0.020 & 0.001 & -0.236 & 0.0  & 0.157 & 
0.001 &  0.959 & 0.002 & 0.002\\
\hline
$\tilde H_1^0$ & -0.310 & -0.002 & 0.655 & 0.0 &  -0.254 & 
0.609 & 0.195 & -0.029 & 0.026\\
\hline
$\tilde H_2^0$ & -0.411 & 0.001 & -0.596 & 0.0 & 0.253 & 
0.606 & -0.197 &  0.059 & -0.054\\
\hline
$\tilde S$ & 0.853 & 0.001 & -0.073 & 0.0 & 0.005 & 
0.512 & -0.002 & -0.046 & 0.039\\
\hline
$\tilde S_1$ & 0.030 &  -0.707 & 0.027 & 0.576  & -0.012 & 
0.003 & 0.009 & 0.305 & -0.270\\
\hline
$\tilde S_2$ & 0.029 & 0.707 & 0.030 & 0.577 & -0.012 & 
 0.003 & 0.009 & 0.304 & -0.269\\
\hline
$\tilde S_3$ & -0.060 & -0.001 & -0.057 & 0.578 & 0.024 & 
 -0.005 & -0.019 & -0.608 & 0.537\\
\hline
\end{tabular}
\end{center}
\end{table}

\renewcommand{\arraystretch}{1.4}
\begin{table}[t]
\caption{The eigenvectors for the CP even 
 Higgses in Model I with $h=0.5$.
\label{tab:eH1}}
\vspace{0.4cm}
\begin{center}
\begin{tabular}{|c|c|c|c|c|c|c|}
\hline        
 Fields & $H_1^0$  & $H_2^0$  & $H_3^0$  & $H_4^0$ 
  & $H_5^0$  & $H_6^0$ \\
\hline
$H_1^{0r}$ &  0.487 & 0.028 & 0.0 & 0.506 & 0.710 & 0.039 \\     
\hline
$H_2^{0r}$ & 0.512 & 0.029 & 0.0 & 0.492 &  -0.699 & -0.080\\     
\hline
$S^{0r}$ & -0.704 & 0.001 &  0.0 & 0.707 & -0.024 & 0.060\\     
\hline
$S_1^{0r}$ & -0.043 & 0.576 & -0.707 & -0.008 & 0.034 & -0.407\\ 
\hline
$S_2^{0r}$ &  -0.042 & 0.576 & 0.708 & -0.007 & 0.034 & -0.406\\     
\hline
$S_3^{0r}$ &  0.037 & 0.579 & 0.0 & -0.036 & -0.068 & 0.811\\ 
\hline
\end{tabular}
\end{center}
\end{table}

\renewcommand{\arraystretch}{1.4}
\begin{table}[t]
\caption{Same as Table~\ref{tab:eH1}, except
$h=0.75$.
\label{tab:eH2}}
\vspace{0.4cm}
\begin{center}
\begin{tabular}{|c|c|c|c|c|c|c|}
\hline        
 Fields & $H_1^0$  & $H_2^0$  & $H_3^0$  & $H_4^0$ 
  & $H_5^0$  & $H_6^0$ \\
\hline
$H_1^{0r}$ & 0.479 &  0.028 & -0.001 & 0.507 & -0.715 &  0.039 \\     
\hline
$H_2^{0r}$ &  0.519 & 0.029 & 0.0 & 0.492 & 0.694 & -0.081\\     
\hline
$S^{0r}$ & -0.704 & 0.001 & 0.0 & 0.707 & 0.033 & 0.060\\     
\hline
$S_1^{0r}$ & -0.043 & 0.576 & -0.706 & -0.008 & -0.034 & -0.407\\ 
\hline
$S_2^{0r}$ & -0.042 & 0.576 & 0.708 & -0.007 & -0.034 & -0.406\\     
\hline
$S_3^{0r}$ &  0.038 & 0.579 & -0.001 & -0.036 & 0.067 & 0.811\\ 
\hline
\end{tabular}
\end{center}
\end{table}

\renewcommand{\arraystretch}{1.4}
\begin{table}[t]
\caption{The eigenvectors for the CP odd 
 Higgses in Model I with $h=0.5$. $G_{1,2}^0$ are mixtures of the
unphysical states absorbed by the $Z$ and \zpr.
\label{tab:eA1}}
\vspace{0.4cm}
\begin{center}
\begin{tabular}{|c|c|c|c|c|c|c|}
\hline        
 Fields & $G_1^0$  & $G_{2}^0$  & $A_1^0$  & $A_2^0$ 
  & $A_3^0$  & $A_4^0$ \\
\hline
$H_1^{0i}$ & -0.665 & 0.212 &  0.0 & -0.322 & -0.013 & 0.639\\     
\hline
$H_2^{0i}$ & 0.670 & -0.256 & 0.0 & -0.314 & -0.013 & 0.623\\     
\hline
$S^{0i}$ & 0.020 & 0.057 & 0.0 & 0.891 &  -0.001 & 0.451\\     
\hline
$S_1^{0i}$ & -0.135 & -0.386 & 0.707 & 0.023 & 0.576 & 0.010\\ 
\hline
$S_2^{0i}$ & -0.135 & -0.384 & -0.707 & 0.023 &  0.578 & 0.010\\     
\hline
$S_3^{0i}$ & 0.269 & 0.768 & 0.001 & -0.059 & 0.578 & 0.008\\ 
\hline
\end{tabular}
\end{center}
\end{table}

\renewcommand{\arraystretch}{1.4}
\begin{table}[t]
\caption{Same as Table~\ref{tab:eA1}, except
$h=0.75$.
\label{tab:eA2}}
\vspace{0.4cm}
\begin{center}
\begin{tabular}{|c|c|c|c|c|c|c|}
\hline        
 Fields & $G_1^0$  & $G_{2}^0$  & $A_1^0$  & $A_2^0$ 
  & $A_3^0$  & $A_4^0$ \\
\hline
$H_1^{0i}$ & -0.696 & -0.013 & 0.0 & -0.322 & -0.013 & 0.641\\     
\hline
$H_2^{0i}$ & 0.718 & -0.026 & 0.0 & -0.312 & -0.013 & 0.622\\     
\hline
$S^{0i}$ & 0.001 & 0.060 & 0.0 & 0.891 & 0.0 & 0.449\\     
\hline
$S_1^{0i}$ & -0.004 & -0.409 & 0.708 & 0.023 & 0.576 & 0.010\\ 
\hline
$S_2^{0i}$ & -0.004 & -0.407 & -0.707 & 0.023 & 0.578 & 0.010\\     
\hline
$S_3^{0i}$ & 0.008 & 0.814 & 0.002 & -0.059 & 0.578 & 0.008\\ 
\hline
\end{tabular}
\end{center}
\end{table}

\end{document}